\documentclass[aps,prl, twocolumn,groupedaddress]{revtex4}

\begin{document}


\title{Comment on ``$^{15}$O($\alpha,\gamma)^{19}$Ne Breakout Reaction and Impact on X-Ray Bursts''}


\author{B. Davids}
\affiliation{TRIUMF, 4004 Wesbrook Mall, Vancouver, BC, Canada V6T 2A3}


\date{\today}



\pacs{25.55.-e, 23.60.+e, 26.30.+k, 26.50.+x}

\maketitle


A recently published letter \cite{tan07} reports a measurement of $\alpha$ decay from states in $^{19}$Ne at excitation energies below 4.5 MeV. The measured $\alpha$ decay branching ratios B$_\alpha$ are used to calculate the astrophysical rate of the $^{15}$O($\alpha,\gamma)^{19}$Ne reaction and to draw conclusions regarding the transition between steady state and unstable nuclear burning on accreting neutron stars. Here I show that the calculated astrophysical reaction rate is based on an unreliable value of B$_\alpha$ for the 4.03 MeV state and point out a serious internal inconsistency in the treatment of low statistics $\alpha$ decay measurements.

The kinematically corrected $\alpha$ particle energy spectra presented in Fig.\ 3 are the basis for the claim that excess counts observed above the background represent $\alpha$ decays from particular states in $^{19}$Ne. The two energy bins between 100 and 400 keV in the upper left panel contain 43 events where the expected background is 34, amounting to an observed excess of 9 events. Attributing these 9 events to $\alpha$ decays from the 4.03 MeV state is questionable in light of the pronounced dip below the background in the two energy bins between 0.8 and 1.1 MeV. These two bins contain a total of 18 events while the estimated background is 27, a deficit of 9 events. The amplitude of this dip relative to its background is greater than the amplitude of the supposed $\alpha$ decay peak relative to its larger background. If this dip is a statistical fluctuation, then there is no reason to suppose the nominal $\alpha$ decay peak is anything other than an statistical fluctuation.

This can also be seen by considering the likelihood function for Poisson-distributed events. Let us suppose that the true value of B$_\alpha$ for the 4.03 MeV state is $\leq10^{-5}$. Then the signal in the counting experiment described in the letter would be 0, because the state would not have been populated often enough for even a single $\alpha$ decay to occur. In this case, with a signal of 0 and an expected background of 34, the probability of observing 43 events is 30\% as large as the probability of observing 34 events. This simple calculation shows why the detection claim made in the letter is unjustifiable. There is no compelling reason to suppose the peak attributed to $\alpha$ decay from the 4.03 MeV state is anything other than an statistical fluctuation.

The letter reports that for the 4.03 MeV state, B$_\alpha = 2.9\pm2.1\times10^{-4}$. It is generally accepted that measurements should have a statistical significance of at least 3$\sigma$ to justify a detection claim. The claim to have detected $\alpha$ decay from a state for the first time with a statistical significance of only 1.4$\sigma$ is both dubious and internally inconsistent. The authors properly dismiss an earlier claim that B$_\alpha = 0.044\pm0.032$ for the 4.38 MeV state \cite{magnus90} on account of poor statistics. Yet their supposed detection of $\alpha$ decay from the 4.03 MeV state has the same 1.4$\sigma$ statistical significance. In a case such as this, only an upper limit can be extracted with any reliability. Applying a Bayesian approach assuming a uniform prior probability density function for nonnegative signals as described by the Particle Data Group \cite{pdg}, the observation of 43 events with an estimated background of 34 implies a 90\% confidence level (CL) upper limit on the signal of 19 events. Since the reported branching ratio is based on a signal of 9 events, the 90\% CL upper limit on B$_\alpha$ is $6.1\times10^{-4}$. This upper limit is considerably larger than the best 90\% CL upper limit to date of $4.3\times10^{-4}$ \cite{davids03a}. Hence the measurement reported in Ref.\ \cite{tan07} represents neither the first definitive detection of $\alpha$ decay from the 4.03 MeV state in $^{19}$Ne nor the best upper limit to date on its $\alpha$ decay branching ratio. Below 0.6 GK, the astrophysical reaction rate is dominated by the resonant contribution of the 4.03 MeV state. Since it is based on an ill-determined value of B$_\alpha$, the reaction rate presented in the letter is not reliably determined from experiment.

Supposed $\alpha$ decays attributed to the 4.14/4.20 MeV doublet are said to be more likely from the 4.14 MeV than the 4.20 MeV state. If this were so, then the reported $\alpha$ decay branching ratio of the 4.14 MeV state along with its measured lifetime \cite{tan05} would imply a reduced $\alpha$ width larger than the Wigner limit, a physical impossibility. This spectrum contains a satellite peak between 1.1 and 1.5 MeV where there should only be background. The statistical significance of the satellite peak clearly exceeds that of the supposed $\alpha$ decay peak. Similarly, the 4.38 MeV state spectrum exhibits a satellite peak more statistically significant than the supposed $\alpha$ decay peak. The background in this measurement is apparently not well described by the estimated random background. These satellite peaks are attributed by the authors to real $\alpha$ decays from higher-lying states in $^{19}$Ne, an explanation that seems problematic. If the satellite peaks in the 4.14/4.20 MeV and 4.38 MeV spectra are due to $\alpha$ decays from higher lying states, then there should be corresponding peaks in the 4.03 MeV spectrum which were not observed. Regardless, this explanation implies that there is an irreducible background from a source not modelled by the authors. But in order to justify the claim that an observed excess with respect to the background represents $\alpha$ decay the background must be described completely and quantitatively, which was not done here.
\bibliography{ne19}

\end{document}